\documentstyle[12pt]{article}
\newcommand{\const}{\mathop{\rm const}\nolimits}

\def\metl{g_{ij}}

\def\metu{g^{ij}}

\def\eml{F_{ij}}
\def\emu{F^{ij}}

\def\be{\begin{equation}}
\def\ee{\end{equation}}
\def\bea{\begin{eqnarray}}
\def\eea{\end{eqnarray}}
\def\tf{\tilde{f}}

\def\tV{\tilde{V}}
\def\tY{\tilde{Y}}
\def\tX{\tilde{X}}
\def\tVp{{\tilde{V}}^{\prime}}

\def\bphi{\bar{\phi}}

\def\tYs{\tilde{Y}}
\def\tYf{{\tilde{Y}}_{\phi}}

\def\ltf{(\log{\tf})_{uv}}

\def\mZ{{\stackrel{\scriptscriptstyle m}{Z}}}

\def\mpsi{\stackrel{\scriptscriptstyle m}{\psi}}

\def\mIpsi{\stackrel{\scriptscriptstyle 1}{\psi}}

\def\mmsum{\sum^{N}_{m=1}}

\begin{document}

\begin{center}
{\bf  A NEW CLASS OF INTEGRABLE MODELS OF
$1+1$ DIMENSIONAL DILATON GRAVITY COUPLED TO SCALAR MATTER}
\end{center}
\vskip 15 pt

\begin{center}
{\sc   {\underline{ A.~T.~FILIPPOV}},
        V.~G.~IVANOV}
\end{center}
\vskip 5 pt

\begin{center}
{\it Bogolubov Laboratory of Theoretical Physics\\
Joint Institute for Nuclear Research, 141980 Dubna, Russia}
\end{center}
\vskip 30 pt

\begin{abstract}
Integrable models of $1+1$ dimensional gravity coupled to
scalar and vector fields are briefly reviewed.
A new class of integrable models with nonminimal coupling
to scalar fields is constructed and discussed.
\end{abstract}

\vskip 20 pt

\centerline{1. INTRODUCTION}
\bigskip

\noindent
Why are the $1+1$ dimensional dilaton-gravity models worth studying?\\
1. They describe some physical objects, e.g. black holes and
cosmological models.           \\
2. Being related to higher-dimensional
gravity and to string theories, they are free of ultra-violet
divergences. We may thus hope to find nonperturbative solutions in the $1+1$
dimensions which give some insight into the problems of higher dimensional
gravity and string theories. \\
3. The $1+1$ dimensional models of
gravity are mathematically interesting --- there exist several classes of
integrable models. Classically, all these integrable models are explicitly
soluble but at the moment only the simplest ones have been exactly quantized.
In this report, we concentrate on classical solutions.

The Einstein-Hilbert theory in the $1+1$ dimensions described by the action
\be
S = \int d^{2}x \, \sqrt{-g} R(g)
\label{f1}
\ee
is known to be trivial because $\sqrt{-g} R(g)$ is a total derivative.
It is easy to prove this in the light-like metric
\be
ds^2=-4f(u,v)\, du \, dv\, ,
\label{f2}
\ee
in which
\be
R=(\log{|f|})_{uv}/f \;\; \mbox{ and }\;\;
\sqrt{-g}R \sim (\log|f|)_{uv}\, .
\label{f3}
\ee

The spherically-symmetric reduction of higher dimensional gravity coupled
to the scalar and vector fields gives a non-trivial $1+1$ dimensional theory.
It is useful to study a more general dilaton-gravity coupled to matter
(we call it DGM; for more details and references see~\cite{Last})
\be
{\cal L} = \sqrt{-g} \, [U R(g) + V + W\metu \varphi_i \varphi_j +
X\eml \emu +
Y + \mmsum \mZ \metu \mpsi_i \mpsi_j] \, ,      \label{f4}
\ee
where $U,\ldots ,\mZ\, ,\; m=1, \ldots ,N$ are functions of $\varphi$,
and $Y$, and $\mZ$ may in addition depend on $\mpsi$.
The subscripts denote partial derivatives
($\varphi_i=\partial_i \varphi$) except when used in
$\metl$ or $\eml$;
$F_{ij}=\partial_i A_j - \partial_j A_i\, ; \; i,j=0,1$ or $i,j=u,v$.

In what follows we use the metric~(\ref{f2}).
If $X=Y=Z=0$, the model is called the dilaton gravity (DG).
Note that $U(\varphi)$ and $W(\varphi)$ may be fixed
to be more or less arbitrary functions.
For example, if $U(\varphi)>0$, we may choose $U(\varphi)=e^{\phi}$.
In what follows we usually choose $U(\varphi)=\phi$.
The potential $W(\varphi)$ can be made an arbitrary function of
$\varphi$ by the Weyl rescaling
\begin{equation}
\begin{array}{c}
\label{f5}
{\cal L}(U, ... , Z)={\bar {\cal L}}(\bar{U}, ... , \bar{Z})\, , \;
g_{ij}=\Omega(\varphi)\bar{g}_{ij}\, ,\; \bar{U}=U\, , \;
\bar{F}^{ij} = \Omega^2 F^{ij}\, ,\;      \\
\\
\bar{Z}=Z\, ,\; \bar{V} = \Omega V\, , \;
\bar{Y} = \Omega Y\, , \; \bar{X} = \Omega^{-1} X \, , \;
\bar{W} = W + U^\prime (\log{\Omega})^\prime\, ,
\end{array}
\end{equation}
where the prime denotes differentiation with respect to $\varphi$.
If we choose
\be
\label{f7}
\Omega=\frac 1{w(\varphi)}\, ,\;\; \mbox{ where }\;\;
w(\varphi)=\exp{\left(
\int{\frac{W(\varphi)}{U'(\varphi)} \, d\varphi} \right)},
\ee
then $\bar{W}=0$. Using the simple transformation
$\bar{w}=\Omega(\varphi) w$, it is easy to find, in addition to
$U(\varphi)$ and $Z(\varphi)$, the following invariants of
the Weyl transformation:
\be
\label{f8}
V/w \equiv \tilde{V}\, ,\;\;
X w = \tilde{X}\, , \;\; X \eml \emu /w\, ,\;\; Y/w \equiv \tilde{Y}\, ,\;\;
f w \equiv \tilde{f}\, .
\ee
Thus, the Lagrangian can be written in the form
\be
{\cal L} = \sqrt{-\bar{g}} \, [U R(\bar{g}) +
\tV + \tX\eml \emu +
\tY + \mmsum \mZ \metu \mpsi_i \mpsi_j] \, .      \label{f9}
\ee

One may further simplify the DGM theory by using the following obvious
integral of motion: $Q=XF_{uv}/{2f}$. This allows one to exclude
$F_{uv}$ and write an effective action for the fields
$\varphi\, , \; f$ and $\psi$ (it is sufficient to add the term
$2Q^2/X$ to the potential $V(\varphi)$).
With all these simplifications and using the parametrization
$U(\varphi)=\varphi^n$ we can write the general DGM in the form
\be
{\cal L} = \sqrt{-g} \, [\varphi^n R(g) + V +
Y + \mmsum \mZ \metu \mpsi_i \mpsi_j] \, .      \label{f10}
\ee

Several subclasses of this general theory are explicitly integrable:
\begin{itemize}
\item The DG model with an arbitrary potential $V(\varphi)$ is integrable.
In fact, it is a topological theory described by a finite number of
degrees of freedom (see e.g.~\cite{Banks},~\cite{Last}).
\item If $\mZ$ are independent of $\varphi$ and $Y=0$,
there exist potentials $V$ for which the model is
integrable (see e.g.~\cite{CGHS},~\cite{Last}).
\item If $V=0\, , \; Y=0$, the equations for the functions
$\varphi\, ,\; \psi \, ,\; f$ are linear but the theory is
not explicitly integrable in general. Nevertheless, there exist
potentials $Z(\varphi)$, for which one may write explicit general solutions.
These new integrable models are introduced in the last section of
this report.
\end{itemize}

Before precisely formulating and proving all these statements,
we present several examples of DGM. The first is the spherically
symmetric gravity in $d=n+2$ dimensions after the dimensional reduction.
It is described by the Lagrangian (see e.g.~\cite{Haj},~\cite{Tada})
\be
{\cal L} = \sqrt{-g} \, [\varphi^n R + n(n-1)/\varphi + 2\Lambda \varphi -
\beta \varphi^n \eml \emu -
\gamma \varphi^n \metu \psi_i \psi_j] \, ,      \label{f11}
\ee
where we have included the cosmological term.
The `induced gravity' of Russo and Tseytlin~\cite{RiT}
can be described by
\be
{\cal L} = \sqrt{-g} \, [\varphi R + 2\Lambda e^{-\lambda \varphi} +
Z \metu \psi_i \psi_j] \, ,      \label{f12}
\ee
The  string-inspired DG, or CGHS~\cite{CGHS} model, is given by
\be
{\cal L} = \sqrt{-g} \, [\varphi R -8\lambda +
Z \metu \psi_i \psi_j] \, .      \label{f13}
\ee
Many other models can be reduced to  these models by redefining
$U(\varphi)$ and by the Weyl rescaling.
There exist more complex models such as the spherically reduced $D=4$
gravity with YM ($SO(3)$)
(so-called Witten model, see e.g.~\cite{Haj})
\be
S=S_{DG} + S_{W}
\ee
\be
S_{W}= \int d^{2}x \, \sqrt{-g}\left\{
\frac 14 \varphi^2 \eml \emu +
(D_i \psi)^{+}(D^i \psi) +
\frac 12  \frac {(e|\psi|^2+P_0)^2}{\varphi^2} \right\}
\label{f14}\, ,
\ee
where $\psi=\psi_1+i\psi_2\, ,\;\;
D_i \psi=\partial_i \psi + ieA_i \psi$, and $A_i$ is an Abelian
vector field.

\vskip 20 pt

\centerline{2. EQUATIONS OF MOTION}
\bigskip

\noindent
It is convenient to write
equations of motion for the action~(\ref{f9})
in the conformal gauge with $U(\varphi)=\phi$ and omitting the $X$-term
\be
\tf (\phi_i / \tf)_i =
\mmsum\mZ (\mpsi_i)^2\;,\;\;(i = u,v) \, ,               \label{5}
\ee
\be
\phi_{uv} + \tf (\tV  + \tYs) = 0 \, ,                    \label{6}
\ee
\be
\ltf +  \tf (\tVp + \tYf) =
\mmsum\mZ{}^{\prime} \mpsi_u \mpsi_v\ \, ,                \label{7}
\ee
\be
(\mZ\mpsi_u)_v + (\mZ\mpsi_v)_u + \tf \tYs_{m} = 0 \, ,
\;\; m=1, \ldots ,N\, .                                  \label{8}
\ee
These equations are not independent.
Thus, eqs.(\ref{5} - \ref{7}) and eq.(\ref{8}) for $m=2,\ldots,N$
imply eq. (\ref{8}) for $m=1$ if
$(\mIpsi_u)^2 + (\mIpsi_v)^2 \neq 0$.
Similarly, eqs.(\ref{5}), (\ref{6}) and eq.(\ref{8}) for $m=1,\ldots,N$
imply eq.(\ref{7}) if $\phi^2_u + \phi^2_v \neq 0$.
The statements are easy to prove by considering
$(\sum_{m=1}^{N} \mZ (\mpsi_u)^2)_v$
and $(\sum_{m=1}^{N} \mZ (\mpsi_v)^2)_u$.
A useful corollary is the following. If $\mZ{}^{\prime} = 0$ and\\
$\tYs=0$,
we only need to solve eqs.(\ref{6}) and (\ref{7}).
The scalar fields $\mpsi$, $m=2,\ldots,N$ are arbitrary free waves.
The scalar field $\mIpsi$ can be derived from eqs.(\ref{5})
while eq.(\ref{8}) for $m=1$ is automatically satisfied.
These simple observations are very useful for analyzing
integrability. Note that (\ref{7}) is often neglected but it is the
key equation in our approach to find integrable DGM models.

The general solution of the models with $Y=Z=0$
can be written in terms of one free field.
The general solution to the constraints~(\ref{5}) can be found
from the obvious relations
$$
\tf/\phi_u=b'(v)\, , \;\; \tf/\phi_v=a'(u)\, ,
$$
where $a(u)$ and $b(v)$ are arbitrary smooth functions. It follows that
\be
\phi=F(r)\, , \;\; \tf=F'(r) a'(u) b'(v)\, , \;\;\;
r \equiv a(u)+b(v)\, .
\label{f18}
\ee
where $F(r)$ is an arbitrary function of the free field $r(u,v)$.
This means that the general dilaton gravity (coupled to Abelian
gauge fields) is a topological theory and can be dimensionally reduced
to the $0+1$ dimensional theory~\cite{Banks},~\cite{Last},~\cite{VDA}.

Now, using~(\ref{f18}) and~(\ref{6}), it is not difficult to show that
$M \equiv F'(r) + N(\phi)$, where $N'(\phi) \equiv \tV(\phi)$, is independent
of $r$. Thus, we find
$$
r=\int 1/(M-N(\phi))\, d\phi
$$
and this defines the general  solution in terms of the free field
$r=a(u)+b(v)$. Note that the integral of motion $M$ can be written
in the covariant form
$$
M=N(\phi)+g^{ij}\phi_i \phi_j /w(\phi)\, .
$$
The equations of motion give $M_i=0$, i.e., $M(u,v)$ is locally conserved.

The solutions essentially defined by functions of one variable $r$
will be called static. More precisely, the static solution can be written as
$\phi=F(r)$, $\tf=E(r) a'(u) b'(v)$, $r=a(u)+b(r)$.
We have shown that all solutions of the DG models are static.
When there is a coupling of the DG to scalar fields,
the general solutions to
equations~(\ref{5})-(\ref{8}) are not static although static
solutions form an interesting subclass of solutions.

\vskip 20 pt

\centerline{3. GENERALIZED CGHS MODEL}
\bigskip

\noindent
For the generalized CGHS model
$Z(\phi)=Z_0=\const$, $Y=X=0$, the function $\tV(\phi)$ is linear:
$\tV(\phi)=g_1 + 2 g_2 \phi \equiv 2 g_2 \bphi$.
The generalized CGHS equations are
\be
\tf\left({\phi_i}/{\tf}\right)_i=Z_0\psi_i^2\, ,\;\; (i=u,v)\, ,
\ee
\be
\bphi_{uv}+ 2 g_2 \bphi \tf=0\, , \label{first_d}
\ee
\be
(\log \tf)_{uv}+2 g_2 \tf =0\, , \label{Liouville}
\ee
\be
\psi_{uv}=0\, .
\ee
The Liouville equation~(\ref{Liouville}) describes the conformal metric.
Its solution can be represented in four forms
which are mutually locally equivalent under coordinate mappings.
The first form of the solution is
\be
\tf_{1}(u,v)= -\frac{a_1'(u) b_1'(v)}{g_2 r_1^2}\, ,\;\;\;
r_1 \equiv a_1(u)+b_1(v)\, .
\label{ssecond}
\ee
The singularity $r_1=0$ separates the first solution~(\ref{ssecond})
into two parts. These parts can be joined into the second solution
\be
a_1=\tan {a_2} \, ,\;\; b_1=-\cot {b_2}\, ,
\label{mapping23}
\ee
\be
\tf_{2}(u,v)= -\frac{a_2'(u) b_2'(v)}{g_2 \cos^2{r_2}}\, ,\;\;\;
r_2 \equiv a_2(u)+b_2(v)\, .
\label{third}
\ee
The first solution can also be transformed into the third and the fourth ones
\be
a_1= e^{2a_{3}}\, ,\;\; b_1=  e^{-2b_{3}}\, ,\;\;\; \mbox{or} \;\;\;
a_1= e^{2a_{4}}\, ,\;\; b_1=- e^{-2b_{4}}\, ,
\label{mapping12}
\ee
\be
\tf_{3}(u,v)= \frac{\displaystyle a_{3}'(u) b_{3}'(v)}
{\displaystyle g_2 \cosh^2{r_3}}\, ,\;\;
r_3 \equiv a_3(u)+b_3(v)\, ,
\label{ffirst1}
\ee
\be
\tf_{4}(u,v)=-\frac{\displaystyle a_4'(u) b_4'(v)}
{\displaystyle g_2 \sinh^2{r_4}}\, , \;\;
r_4 \equiv a_4(u)+b_4(v)\, .
\label{ffirst2}
\ee
The solutions $\tf_{3}$ and $\tf_{4}$ are two parts of a single
solution describing two nearby maps of the manifold.
It is possible to join them by a coordinate substitution.

It is easy to derive a general nonstatic solution for the dilaton
and the scalar field. For the first form of the solution
(see~(\ref{ssecond})), it is~\cite{Last}
\be
\label{dilaton}
\bphi=(A_1'(a_1)+B_1'(b_1))-2(A_1(a_1)+B_1(b_1))/r_1\, ,
\ee
\be
\frac{\partial \psi}{\partial a_1} \equiv
\psi_{a_1}=\sqrt{A_1'''/Z_0}\, ,\;\;\;
\frac{\partial \psi}{\partial b_1} \equiv
\psi_{b_1} \, = \, \sqrt{B_1'''/Z_0}\, ,
\ee
where $A_1(a_1)$, $B_1(b_1)$ are arbitrary functions.
The dilaton and the scalar field for the
metrics~(\ref{third},~\ref{ffirst1},~\ref{ffirst2})
can be reproduced by coordinate mappings~(\ref{mapping12},~\ref{mapping23})
in the solution~(\ref{dilaton})
\be
\bphi= (A_2'(a_2)+B_2'(b_2))+2(A_2(a_2)+B_2(b_2))
\tan(r_2)\, ,
\label{dilaton2}
\ee
\be
Z_0 \psi_{a_2}^2 = A_2'''(a_2)+4A_2'(a_2)\, , \;\;
Z_0 \psi_{b_2}^2 = B_2'''(b_2)+4B_2'(b_2)\, ,
\ee
\be
\bphi= (A_3'(a_3)+B_3'(b_3))-2(A_3(a_3)+B_3(b_3))\tanh{r_3}\, ,
\label{dilaton3}
\ee
\be
\bphi= (A_4'(a_4)+B_4'(b_4))-2(A_4(a_4)+B_4(b_4))\coth{r_4}\, ,
\label{dilaton4}
\ee
\be
Z_0 \psi_{a_m}^2=  A_m'''(a_m)-4A_m'(a_m)\, ,\;\;
Z_0 \psi_{b_m}^2=  B_m'''(b_m)-4B_m'(b_m)\, ,\;\; m=3,4\, .
\ee
The functions $A_m(a_m)$ (as well as $B_m(b_m)$)
corresponding to different values of $m=1,...,4$ are different
but related to one another.

The most general integrable model with $Z=Z_0$ and
$\tV=g_{+}e^{g\phi}+g_{-}e^{-g\phi}$ has been introduced
in~\cite{Last}. Its solution is essentially reduced
to solving two Liouville equations.
Here, we introduce a new class of integrable DGM models
which are reduced to linear, explicitly soluble equations.

\vskip 20 pt

\centerline{4. NONREFLECTING POTENTIALS}
\bigskip

\noindent
Let us study the DG model with $V=0$ and
minimal coupling to one scalar field.
Equation~(\ref{8}) becomes $\phi_{uv}=0$, i.e., $\phi=a(u)+b(v)$
where $a(u)$ and $b(v)$ are arbitrary functions. In the coordinates
$(a,\, b)$ the remaining equations~(\ref{5})-(\ref{7}) are
\be
(\log \tf)_i = - Z(\phi) (\psi_i)^2\; ,\;\;(i =a,b) \, ,         \label{m5}
\ee
\be
(\log \tf)_{ab} = Z'(\phi) \psi_a \psi_b \, ,       \label{m7}
\ee
\be
(Z(\phi)\psi_a)_b + (Z(\phi)\psi_b)_a = 0 \, .       \label{m8}
\ee
Equation~(\ref{m7}), depending on the other equations, can be omitted.
For some special forms of the function $Z(\phi)$,
eq.~(\ref{m8}) can be solved exactly.
Then, the metric $\tf$ can be found from eq.~(\ref{m5}).
If $Z(\phi)$ has no zeroes (except boundaries),
we can write $Z(\phi)=\pm \alpha^{2}(\phi)$.
Let us make the substitution
$\psi \equiv \chi(r,t)/ \alpha\, ,\;\;
\alpha \equiv  \alpha(r) \equiv \alpha(\phi)$
and rewrite eq.~(\ref{m8})
in the coordinates $(r,t)$ ($r=a+b \equiv \phi\, ,\; t=a-b$):
\be
\chi''-{\ddot \chi}+G(r)\chi=0\, ,  \label{first}
\ee
where $G(r) \equiv - \alpha''/\alpha$.

For a few classes of nonreflecting potentials,
we can solve eq.~(\ref{first}) in terms of a free wave.
For the nonreflecting potential (see e.g.~\cite{Leznov})
\be
G=G_1(r_1) \equiv -l(l+1)/r_1^2\, , \;\; r_1=a_1+b_1\equiv \phi \, ,
\;\;l=1, 2 , ...
\label{standard}
\ee
we will search for a solution of~(\ref{first})
as the following sum:
\be
\chi=\sum_{k=0}^{n} 2^{n-k} C_{k}(r_1) \Psi^{(n-k)}_{r_1}=
\sum_{k=0}^{n} C_{k}(r) (A^{(n-k)}(a_1)+B^{(n-k)}(b_1))\, ,
\label{sum}
\ee
$$
4\Psi_{a_1b_1}=\Psi''-{\ddot \Psi}=0\, , \;\;\;
\Psi=A(a_1)+B(b_1)\, ,
$$
$$
\Psi^{(m)}_{r_1} \equiv \partial^{m}_{r_1}\Psi(r_1, t_1)\, ,\;\;\;
A^{(m)}(a_1) \equiv \partial^{m}_{a_1}A(a_1)\, ,\;\;\;
B^{(m)}(b_1) \equiv \partial^{m}_{b_1}B(b_1)\, ,
$$
where $A(a_1), \,B(b_1)$ are arbitrary functions,
and $n$ is a positive integer.
After substitution of the sum~(\ref{sum}) into eq.~(\ref{first}),
we obtain the chain of ordinary differential equations for
the functions $C_k(r_1)$
\begin{equation}
\begin{array}{c}
C_{0}'=0\, ,\\
C_{1}'+ G_1(r_1)C_{0}=0\, ,\\
\dots\\
C_{k+1}'+C_{k}''+G_1(r_1)C_k=0\, ,\\
\dots
\label{chain}
\end{array}
\end{equation}
The system can easily be integrated from the first to the last equation.
After fitting the integration constants the chain terminates at $n=l$
\be
C_{n}''+G_1(r_1)C_n=0\, , \;\;\; C_{n+1}=0.
\ee
The solution in terms of the free wave $\Psi(r_1, t_1)$ is
\be
\chi=\sum_{k=0}^{l} C_k(r)
\Psi^{(l-k)}_{r_1} 2^{l-k}\, , \;\;
C_k(r_1)=\frac{(l+k)!}{k! (l-k)!} \left( \frac{-1}{r_1} \right)^k \, .
\label{sum1}
\ee

Using the solution~(\ref{sum1}), we can easily write down the solution for
other nonreflecting potentials\footnote{
All such nonreflecting potentials are the static solutions to the
Liouville equation~(\ref{Liouville}) with $g_2=1/(l(l+1))$.}
\be
G_2(r_2)\equiv -\frac{\lambda^2 l(l+1)}{\cos^2(\lambda r_2)}\, , \;\;
G_3(r_3)\equiv \frac{\lambda^2 l(l+1)}{\cosh^2(\lambda r_3)}\, , \;\;
G_4(r_4)\equiv -\frac{\lambda^2 l(l+1)}{\sinh^2(\lambda r_4)}\, .
\label{potentials}
\ee
One can put $\lambda=1$ by rescaling.
Here, we suppose $\phi = r_m \equiv a_m + b_m\, , \;\; m=2,3,4$
respectively.
Primes and dots in eq.~(\ref{first}) with the
potentials~(\ref{potentials}) denote the derivatives with respect to
the new variables $r_m$ and
$t_m\, ,\;\; m=2,3,4$.

The solutions to eq.~(\ref{first}) with the
potentials~(\ref{potentials}) can be found directly
by eqs.~(\ref{chain}) with $n=l$.
A simpler way is
the substitutions~(\ref{mapping23}), (\ref{mapping12}).
Under these substitutions,
equation~(\ref{first}) with the potential $G_1(r_1)$
transforms into itself but with the new potentials
$G_2(r_2)$, $G_3(r_3)$, or $G_4(r_4)$.
After these substitutions, the sum~(\ref{sum1}) (the solution
of the eq.~(\ref{first}) with the potential $G_1(r_1)$)
with $r_1=r_1(a_m,b_m) \equiv a_1(a_m)+b_1(b_m)$
gives us solutions of eq.~(\ref{first}) with the potentials
$G_m(r_m)$, $m=2,3,4$.
The functions
$C_k(r_1)=C_k(a_1(a_m), b_1(b_m))=C_k(r_m,t_m)\, ,\;\; m=2,3,4$
are no more static.

Finally, to find all the functions $\alpha(\phi)$,
which give us the potential $G(\phi)$ in the
forms~(\ref{standard}) and (\ref{potentials}),
one has to solve the equation
\be
\alpha''(\phi)+G(\phi)\alpha(\phi)=0\, , \label{second}
\ee
which is equivalent to the static form of eq.~(\ref{first}).
The solution of eq.~(\ref{second}) with the
potential $G_1(\phi)$ can easily be found:
\be
\alpha=K \phi^{l+1}+L \phi^{-l}\, , \;\;\;
Z=\alpha^2=(K \phi^{l+1}+L \phi^{-l})^2\, , \;\;\;
K,\, L = \const\, .
\ee
The solutions to eq.~(\ref{second}) with the
potentials~(\ref{potentials}) are more complicated
\be
\begin{array}{c}
\alpha_m(\phi)=K_m\alpha_{m1}(\phi)+L_m\alpha_{m2}(\phi)\, ,\;\;\;
K_m,L_m=\const\, , \;\;\; m=2,3,4\\
\alpha_{21}(\phi)=P_l(i \tan(\phi))\, ,\;\;
\alpha_{31}(\phi)=P_l(\coth(\phi))\, ,\;\;
\alpha_{41}(\phi)=P_l(\tanh(\phi))\, ,\\
\alpha_{m2}(\phi)=\alpha_{m1}(\phi) \int 1/\alpha_{m1}^2(\phi) \, d\phi\, ,
\end{array}
\ee
where $i$ is the imaginary unit and $P_l(x)$ is the Legendre polynomial.

By the way, eq.~(\ref{first_d}) for the generalized
CGHS model (with the conformal metric in the static forms)
and eq.~(\ref{first}) for the nonreflecting model with $l=1$
are identical.
Therefore, the solutions to eq.~(\ref{first}) with
the potentials~(\ref{standard},~\ref{potentials}) with $l=1$ are already
written in
eqs.~(\ref{dilaton},~\ref{dilaton2},~\ref{dilaton3},~\ref{dilaton4}).
We must only substitute $g_2 \to 1$, $\tf(r_m) \to G(r_m)$,
and $\bphi \to \chi$.

\vskip 30pt

This investigation was partially supported
by the Russian Foundation for Basic Research (project 97-01-01041),
and by INTAS (project 93-127-ext).

\end{document}